\documentclass[12pt]{iopart}
\usepackage{graphicx}
\usepackage{iopams}
\usepackage{epsfig}

\begin{document}

\title{Collective excitations of atoms and field modes in coupled cavities}
\author{Nicolae A. Enaki, Sergiu Bazgan}

\address{Institute of Applied Physics, Academy of Sciences of Moldova,
Academiei Str. 5, Chisinau MD 2028, Republic of Moldova}
\ead{enakinicolae@yahoo.com}
\begin{abstract}
The exact solution of the system consisted from two or three q-bits doped in coupled cavities is discussed. The problem of indistinguishable between the excited radiators and photons is analyzed using the intrinsic symmetry of the system. It is demonstrated that the solution is drastically simplified when the radiators and photons are considered as a new polariton excitations. The exact solution of Schrodinger equation is obtained for single and two excitations in each cavity taking into consideration the indistinguishable principle. This approach opens new possibilities in the interpretation of quantum entangled states in comparison with the traditional distinctive situation (see for example \cite{NM-2001,EB-2013}) due to the decreasing of the number of degrees of freedoms in the system. Considering that the energies of coupling between the radiators and photons is larger than the coupling with external vacuum field, we have found the master equation for the dumping of collective excitations of the system of coupled radiators through the cavity fields. The time-dependence of population for new dressed quasi-levels of energy is obtained solving analytically and numerically the master equation.
\end{abstract}


\section{Introduction}

In recent years the cooperative interaction of $N$ two level radiators with cavity electromagnetic field (EMF) was in the attention of many experimental and theoretical researches \cite{DLCZ}. This it is connected with the big application of two-level system as a q-bit in quantum processing of information. In many cases, distinguished ensembles of q-bits are used for the realization of quantum registers \cite{FCR,KBBC}. According to the principle of indistinguishable between the radiators \cite{D}, $2^{N}$ states of $N$ two-level atoms can be reduced to $N+1$ states in the processes of coherent excitation. As a particular case, the cooperative effect between two undistinguished radiators occurs, due to the fact that two states of the single excitation ( the atom $A$ in an excited state and $B$ in the ground and respectively atom $A$ in the ground state and $B$ in the excited state) are considered as a same collective state. It is not difficult to observe that this number of collective excitation drastically is reduced with increasing the number of atoms.

It is attractive from physical points of view to apply this principle to the atoms placed in coupled cavities. Combination of atom-cavity physics and photon systems offers new opportunities in this field for a better device functionality and for probing of emulators of condensed-matter systems. In Ref. \cite{QMSGH-2009} a single-polariton approximation was proposed for the study of the periodical photonic systems. Within this approximation the authors of this paper have applied Bloch states to the uniformly tuned Jaynes-Cummings-Hubbard model to analytically determine the energy-band structure. In the work \cite{ASB-2007} it was studied the photon-blockade-induced Mott transitions and XY spin models in coupled cavity arrays. It was found that a range of many-body effects such as a Mott transition for polaritons obeying mixed statistics could be observed in the optical systems of individual addressable coupled cavity arrays interacting with two-level systems. Exploited first in systems like photons, atoms, or ions, the entanglement is becoming more and more attainable in condensed matter physics.
As follows from the distinctive description of doped cavities \cite {GMMTG-2011},\cite{EB-2013}, the analytical descriptions of the quantum systems become complicated due to the increasing number of degrees of freedom of such systems by increasing the number of cavities in the system. In this case the number of degrees of freedoms is connected with the numbers of coupled quantum oscillators: atoms and cavity-modes. Considering that the excitations of atoms and cavity modes become indistinguishable, we can use the symmetry transformation of such system considering the invariance of the states after the actions of the operation of corresponded symmetry group.

For example, the wave function of the single excitation of the radiator or mode inside of one of the three coupled cavities must remain in the same quantum state after the rotation around the axis of symmetry with angle $\mathbf{2}\pi/3$ and $\mathbf{4}\pi/3$. As it is observed, the number of degrees of freedom of the system formed from three coupled cavities is reduced as in the Dicke problem for three radiators situated in the volume with the dimension less than the radiation wavelength. In sec.\ref{Sec:2} we have proposed to use the symmetry principle for reducing the number of degrees of freedom of two and three coupled cavities. Due to this, we have obtained the exact solution for the Schrodinger equation in interaction picture. Introducing the losses from the cavities, the master equation for undistinguished excitation of two and more cavities is studied in the sec.\ref{Sec:3}.

\section{Collective excitation of coupled cavities}
\label{Sec:2}
In this section we apply the symmetrical propriety of undistinguished atoms and photons packed in coupled cavities. For comparison we consider the system of atoms situated at a distance less than the emission wavelength as in the Dicke model \cite{D}. For simplicity we consider the system of two atoms. These distinguished atoms are characterized by four states: $|e_{1}e_{2}\rangle$; $|e_{1}g_{2}\rangle$; $|g_{1}e_{2}\rangle$ and $|g_{1}g_{2}\rangle$. Here $e_{l}$ and $g_{l}$ are excited and ground states of $l$ atom, $l=1,2$. Considering that one of the atoms from the ensemble is in excited state we construct the collective excitation state  $|eg\rangle=\frac{1}{\sqrt{2}}\left({|e_{1}g_{2}\rangle}+{|e_{2}g_{1}\rangle}\right)$ which together with other two states $|ee\rangle$ and $|gg\rangle$ form three atomic collective states. It is important to emphasize, that the rotation of this ensemble at angle $\pi$ these collective states remain the same. For three atoms situated in vertexes of equidistant triangle, we can obtain four collective states $|ggg\rangle$ , $|gge\rangle$, $|gee\rangle$ and $|eee\rangle$ from $8$ distinguished $|g_{1}g_{2}g_{3}\rangle$, $|g_{1}g_{2}e_{3}\rangle$, $|g_{1}g_{3}e_{2} \rangle$, $|g_{3}g_{2}e_{1}\rangle$, $|e_{1}g_{2}e_{3}\rangle$, $|g_{1}e_{2}e_{3}\rangle$, $|e_{1}e_{2}g_{3}\rangle$ and $|e_{1}e_{2}e_{3}\rangle$ . As follows from this description of two collective states $|gge\rangle$, and $|gee\rangle$, we observe that they can be regarded as a superposition of three distinguished states ${|g_{1}g_{2}e_{3}\rangle+|g_{1}g_{3}e_{2}\rangle+|g_{3}g_{2}e_{1}\rangle}/\sqrt{3}$, ${|e_{1}g_{2}e_{3}\rangle+|g_{1}e_{2}e_{3}\rangle+|e_{1}e_{2}g_{3}\rangle}/\sqrt{3}$, and are invariant under the angle rotations $2\pi/3$ and $4\pi/3$.

\begin{figure}[tbp]
\begin{centering}
\resizebox{0.28 \textwidth}{!}{\includegraphics{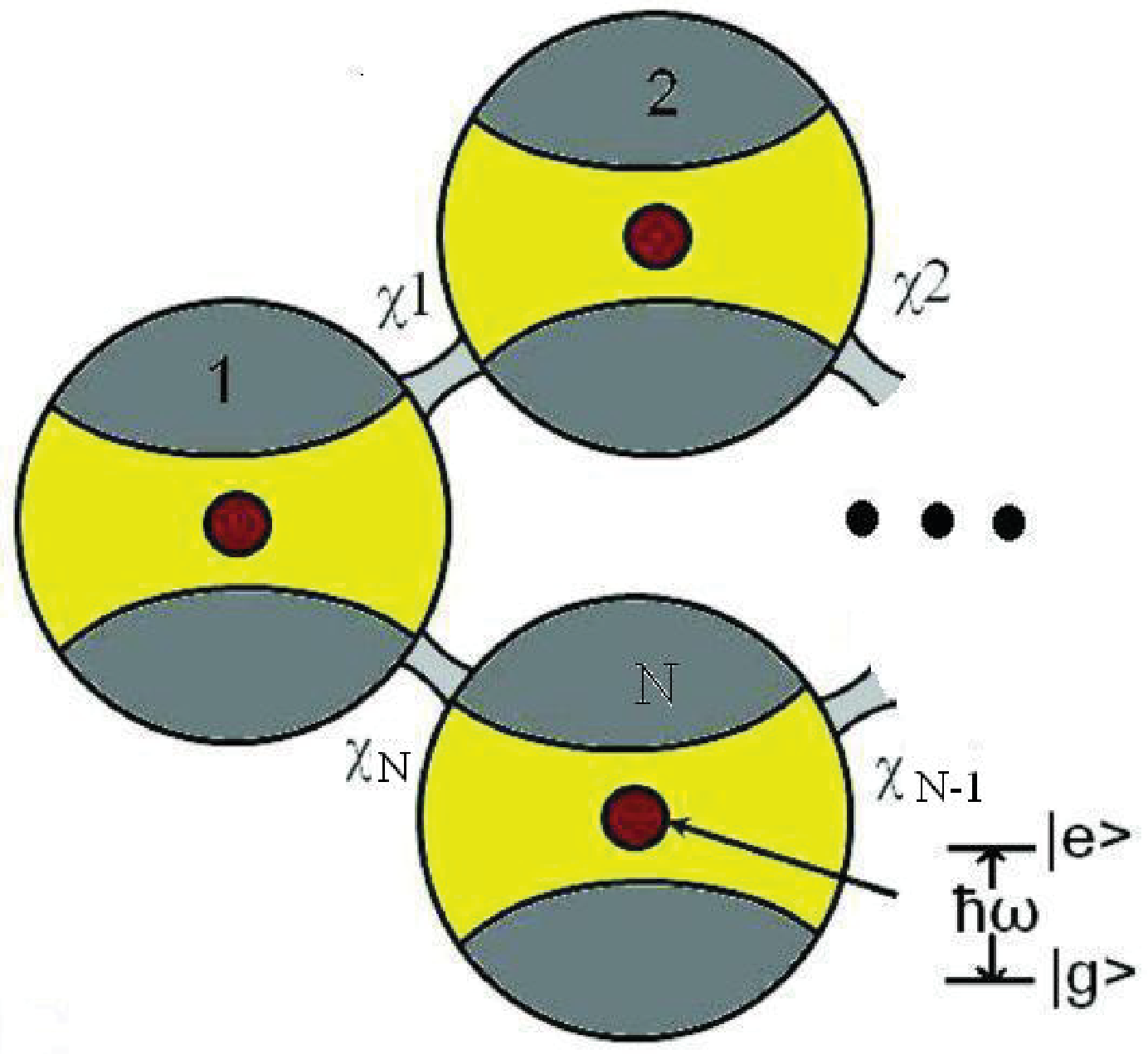}}
\resizebox{0.22 \textwidth}{!}{\includegraphics{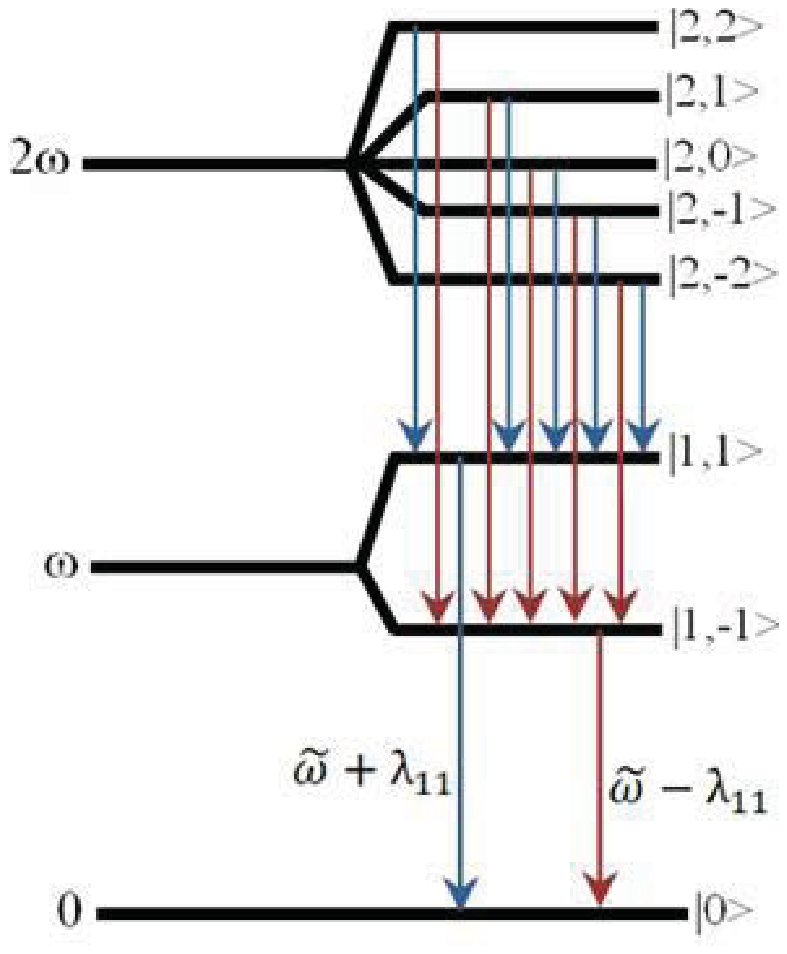}}
\caption{N radiators in interaction with cavity modes of coupled cavities and collective excitations of cavity modes and atoms}
\end{centering}
\label{figure}
\end{figure}

Let us return to the coupled ensemble of n doped cavities. In comparison with the above situation we have two localized excitations in each cavity. First of them is connected with atomic excitation and second with mode excitation in each cavity. In order to apply the proposed symmetry we introduce the Hamiltonian of such coupled cavities in the interaction with the external electromagnetic field. Indeed considering the $n$ single mode coupled cavities doped with radiators as this is represented in the Fig. \ref{figure}), we can introduce the following \ Hamiltonian $H=H_{0}+H_{I},$ in which the free part describes the energies of the atomic inversion, single-mode EMF of each cavity
\begin{equation}
H_{0}=\hbar \omega \sum\limits_{i=1}^{n}\left( R_{zi}+a_{i}^{\dagger}a_{i}\right)  \label{eq:H0}
\end{equation}
The interaction Hamiltonian contains the following couplings: between the atom and the mode of each cavity, between the cavities
\begin{equation}
\fl H_{I}=\hbar
g\sum\limits_{i=1}^{n}(R_{i}^{+}a_{i}+R_{i}^{-}a_{i}^{\dagger })+\hbar \chi
\sum\limits_{i=1}^{n}\sum\limits_{j=1,j\neq i}^{n}\left( a_{i}a_{j}^{\dagger
}+H.c.\right) +H.c.  \label{eq:HI}
\end{equation}
Here $\omega $ is the resonant frequency between two level radiators and modes of the electromagnetic field, $\hat{a}_{i}^{\dagger }$and $a_{I}$ are the creation and annihilation boson operators of EMF , $R_{i}^{+}$, $ R_{i}^{-}$ and $R_{zi}$ are the excitation , de-excitation and inversion atomic operators for the $i$-cavity. These operators satisfy the commutation relation $\left[ R_{i}^{+},R_{j}^{-}\right] =2\delta _{i,j}R_{zi} $, $\left[ R_{zi},R_{j}^{\pm }\right] =\pm R_{i}^{\pm }$. The coupling between the cavities is achieved by the second term of the interaction Hamiltonian described by the exchange parameter $\chi $.

Following the idea, that the system formed from $n$ cavities must remain in the same state after the application of rotation symmetry to regular polygon in the vertexes of which are situated the coupled cavities, we introduce the wave functions of the Hamiltonian (\ref{eq:H0}). Let us neglect in first approximation the coupling of the cavities with external EMF. Solving the problem in higher Q cavity limit we will introduce the losses from the cavities in the next section. Let us below apply these symmetry proprieties of two and three cavities.

\textit{Two cavities situation .} In the good cavity limit, for single excitation in both cavities it is easy to observe $4$ wave functions for disinfested atoms and cavity modes described by free Hamiltonian (\ref{eq:H0}): $|e,0\rangle _{1}|g,0\rangle _{2}$, $|g,0\rangle _{1}|e,0\rangle _{2}$, $|g,0\rangle _{1}|e,1\rangle _{2}$, $|e,0\rangle _{1}|1,0\rangle $. For undistinguished situation we obtain two wave functions
\begin{eqnarray}
\left\vert \psi _{11}\right\rangle &=&\frac{1}{\sqrt{2}}\{\left\vert e,0\right\rangle _{1}\left\vert g,0\right\rangle _{2}+e^{i\phi }\left\vert
g,0\right\rangle _{1}\left\vert e,0\right\rangle _{2}\};  \nonumber \\
\left\vert \psi _{12}\right\rangle &=&\frac{1}{\sqrt{2}}\{\left\vert g,1\right\rangle _{1}\left\vert g,0\right\rangle _{2}+e^{i\phi }\left\vert
g,0\right\rangle _{1}\left\vert g,1\right\rangle _{2}\};  \label{eq:WF12}
\end{eqnarray}
For two excitations in two cavities instead of the $8$ wave functions $|g,1\rangle _{1}|g,1\rangle _{2}$, $|g,2\rangle _{1}|g,0\rangle _{2}$, $|g,2\rangle _{1}|g,0\rangle $, $|e,0\rangle _{1}|e,0\rangle _{2}$; $|e,1\rangle _{1}|g,0\rangle _{2}$; $|e,1\rangle _{1}|g,0\rangle _{2}$; $|e,1\rangle _{1}|g,0\rangle _{2}$; $|g,0\rangle _{1}|e,1\rangle _{2}$ we obtain $5$ undistinguished
\begin{eqnarray}
\left\vert \psi _{22}\right\rangle &=&\frac{1}{\sqrt{2}}\left( \left\vert
e,1\right\rangle _{1}\left\vert g,0\right\rangle _{2}+e^{i\phi }\left\vert
g,0\right\rangle _{1}\left\vert e,1\right\rangle _{2}\right) ;  \nonumber \\
\left\vert \psi _{23}\right\rangle &=&\frac{1}{\sqrt{2}}\left( \left\vert
e,0\right\rangle _{1}\left\vert g,1\right\rangle _{2}+e^{i\phi }\left\vert
g,1\right\rangle _{1}\left\vert e,0\right\rangle _{2}\right) ;  \nonumber \\
\left\vert \psi _{24}\right\rangle &=&\frac{1}{\sqrt{2}}[\left\vert
g,2\right\rangle _{1}\left\vert g,0\right\rangle _{2}+e^{i\phi }\left\vert
g,0\right\rangle _{1}\left\vert g,2\right\rangle _{2}];  \nonumber \\
\left\vert \psi _{21}\right\rangle &=&\left\vert e,0\right\rangle
_{1}\left\vert e,0\right\rangle _{2};\ \ \left\vert \psi _{25}\right\rangle
=\left\vert g,1\right\rangle _{1}\left\vert g,1\right\rangle _{2};
\label{eq:WF22}
\end{eqnarray}

Here $e^{i\phi }=1(\phi =0)$ for symmetrical functions and $-1(\phi =\pi,-\pi )$ for anti-symmetrical case, $\left\vert \psi _{\alpha
i}\right\rangle =\left\vert a_{1},n_{1}\right\rangle _{1}\left\vert a_{2},n_{2}\right\rangle _{2}$ is the state of $\alpha $ \ degenerate level
of energy, where $\alpha =n_{1}+n_{2}+\delta _{e,a_{1}}+\delta _{e,a_{2}}$ is the number of excitations in the system, $a_{i}$ and $n_{i}$ indicate the atomic and field excitations in the $i$ cavity respectively. This wave functions (\ref{eq:WF12}) and (\ref{eq:WF22}) are invariant under the rotation of the system with angle $\pi $. For the rotation of the system of two cavities at angle $\pi $ we must replace the indexes of the cavities between them $1\leftrightarrow 2$. It is easy to observe that for one quanta of energy we have and other superposition functions for two coupled cavities (one photon in first and the atom-excitation in second and vice versa). But these wave functions represent a superposition of collective excitations of atomic and field states of both cavities.

Let consider that the degenerate states for single (\ref{eq:WF12}) and two excitations (\ref{eq:WF22}) are weak split by the interaction Hamiltonian (\ref{eq:HI}), so that this splitting is less than the energy distance between first and second excited states of the cavities. Representing the non-stationary wave function through the superpositions of the single (\ref{eq:WF12}) or two excitations (\ref{eq:WF22}) states $\left\vert \psi(t)\right\rangle $=$\Sigma _{j}\exp [-i\lambda _{j}t]\left\vert \psi_{j}\right\rangle $ , we can find the new dressed states of the system solving the stationary Schrodinger equation in the interaction picture $H_{I}|\psi _{j}\rangle =\lambda _{j}|\psi _{j}\rangle $. The wave functions for single collective excitations are $|\psi _{1}(t)\rangle_{i}=\sum_{j=1,2}c_{i,j}^{(1)}\exp [-i\lambda _{1,i}t]|\psi _{1j}\rangle $. Here $i$ indicate the new energy level, the eigenvalues are
\begin{equation}
\lambda _{1,1(-1)}=\frac{1}{2}\left( e^{i\phi }\chi \pm \sqrt{4g^{2}+\chi
^{2}}\right) ;  \label{eq:L1}
\end{equation}
and the superposition coefficients
\begin{equation}
\fl c_{i,1}^{\left( 1\right) }=\sqrt{g^{2}/(g^{2}+\lambda _{1,i}^{2})},\ \
c_{i,2}^{\left( 1\right) }=\sqrt{\lambda _{1,i}^{2}/(g^{2}+\lambda
_{1,i}^{2})},\ \ \ for\ i=-1,\ 1;  \label{eq:C1}
\end{equation}
are labeled so that the first index indicates the new level of energy, the index in parentheses indicate the number of excitations from the cavities.

For two excitations in the system we have the wave function $|\psi _{2}(t)\rangle _{i}=\sum_{j=1}^{5}c_{i,j}^{(2)}\exp [-i\lambda _{2,i}t]|\psi
_{2j}\rangle $. Here the solutions of characteristic equations are
\begin{eqnarray}
\lambda _{2,1(-1)} &=&\pm \frac{1}{\sqrt{2}}\sqrt{5g^{2}+3\chi ^{2}-\sqrt{%
\left( 3g^{2}+\chi ^{2}\right) ^{2}+12(1+e^{i\phi })g^{2}\chi ^{2}}},
\nonumber \\
\lambda _{2,2(-2)} &=&\pm \frac{1}{\sqrt{2}}\sqrt{5g^{2}+3\chi ^{2}+\sqrt{%
\left( 3g^{2}+\chi ^{2}\right) ^{2}+24g^{2}\chi ^{2}}},\ \ \lambda _{2,0}=0,
\label{eq:L2}
\end{eqnarray}
and the new coefficients for two collective excitations can be found from the normalization condition
\begin{eqnarray}
\fl c_{i,2}^{\left( 2\right) } &=&\frac{\chi }{\sqrt{2}g}\left( 1-\frac{%
3g^{2}}{2\chi ^{2}+g^{2}-\lambda _{2,i}^{2}}\right) c_{i,1}^{\left( 2\right)
};\ \ c_{i,3}^{\left( 2\right) }=\frac{\lambda _{2,i}}{\sqrt{2}g}%
c_{i,1}^{\left( 2\right) };  \nonumber \\
\fl c_{i,4}^{\left( 2\right) } &=&\frac{-3\lambda \chi c_{i,1}^{\left(
2\right) }}{\sqrt{2}\left( 2\chi ^{2}+g^{2}-\lambda _{2,i}^{2}\right) };\ \
c_{i,5}^{\left( 2\right) }=\left( 1-\frac{3\chi ^{2}}{2\chi
^{2}+g^{2}-\lambda _{2,i}^{2}}\right) c_{i,1}^{\left( 2\right) };\ \ for\ \
i=-2...2;  \nonumber \\
\fl c_{i,1}^{\left( 2\right) } &=&\frac{\sqrt{2}g\sqrt{\left( 2\chi
^{2}+g^{2}-\lambda _{2,i}^{2}\right) ^{2}}}{\sqrt{\lambda _{2,i}^{6}+\lambda
_{2,i}^{4}(2g^{2}-3\chi ^{2})+\lambda _{2,i}^{2}g^{2}(13\chi
^{2}-7g^{2})+4\left( \chi ^{6}+g^{6}\right) +8g^{4}\chi ^{2}+2g^{2}\chi ^{4}}
}  \label{eq:C2}
\end{eqnarray}
Here indexes of $c_{n,i}^{\left( m\right) }$: $m$, $n$ and $i$ indicate the number of quanta of energy in the system, the new quasi-energy level number corresponding to the energy $\hbar \left( m\omega +\lambda _{n}\right) $ and the state $\left\vert \psi _{mi}\right\rangle $ used in superpositions. The expressions (\ref{eq:L1}), \ref{eq:L2}) and (\ref{eq:C1}) (\ref{eq:C2}) represent the frequencies of new quasi-levels of energy and wave function coefficients for single and two excitations.

\textit{Three cavity case.} For three coupled doped cavities we can obtain in the similar way the wave function of the single excitation. This
methodology can be extended for larger numbers of excitations representing the eigenstates and wave functions numerically. Taking this into account it is studied only the Master equation for one excitation in system, for which is possible to obtain analytical solutions. The degenerate states for one excitation are
\begin{eqnarray}
\fl\left\vert \psi_{11}\right\rangle & = & \frac{1}{\sqrt{3}}%
\left(\left|e,0\right\rangle {}_{1}\left|g,0\right\rangle
{}_{2}\left|g,0\right\rangle _{3}+\left|g,0\right\rangle
{}_{1}\left|e,0\right\rangle {}_{2}\left|g,0\right\rangle
_{3}+\left|g,0\right\rangle {}_{1}\left|g,0\right\rangle
{}_{2},\left|e,0\right\rangle {}_{3}\right)  \nonumber \\
\fl\left\vert \psi_{12}\right\rangle & = & \frac{1}{\sqrt{3}}%
\left(\left|g,1\right\rangle {}_{1}\left|g,0\right\rangle
{}_{2}\left|g,0\right\rangle _{3}+\left|g,0\right\rangle
{}_{1}\left|g,1\right\rangle {}_{2}\left|g,0\right\rangle
_{3}+\left|g,0\right\rangle {}_{1}\left|g,0\right\rangle
{}_{2},\left|g,1\right\rangle {}_{3}\right)  \label{eq:3c}
\end{eqnarray}
We observe and other superpositions functions of the single excitation in the three cavities. These are the superposition with one photon in one of them and atomic excitation in another. Similarly, it is founded the eigenstates and eigenvalues of interaction Hamiltonian
\begin{eqnarray}
\lambda_{1,1(-1)} & = & \chi\pm\sqrt{\chi^{2}+g^{2}},  \nonumber \\
c_{i,1} & = & \sqrt{g^{2}/(g^{2}+\lambda_{1,i}^{2})};\ \ c_{i,2}=\sqrt{%
\chi^{2}/(g^{2}+\lambda_{1,i}^{2})}\ \ for\ \ i=-1,\ 1;  \label{eq:3cv}
\end{eqnarray}
So, the analyzed interaction allows us to get two new quasi-energy levels. For two excitations we observe five collective states
\begin{eqnarray*}
\fl\left\vert \psi _{21}\right\rangle =\frac{1}{\sqrt{3}}\left( \left\vert
e,0\right\rangle _{1}\left\vert e,0\right\rangle _{2}\left\vert
g,0\right\rangle _{3}+\left\vert e,0\right\rangle _{1}\left\vert
g,0\right\rangle _{2}\left\vert e,0\right\rangle _{3}+\left\vert
g,0\right\rangle _{1}\left\vert e,0\right\rangle _{2}\left\vert
e,0\right\rangle _{3}\right) \\
\fl\left\vert \psi _{22}\right\rangle =\frac{1}{\sqrt{6}}\{\left\vert
e,0\right\rangle _{1}\left( \left\vert g,1\right\rangle _{2}\left\vert
g,0\right\rangle _{3}+\left\vert g,0\right\rangle _{2}\left\vert
g,1\right\rangle _{3}\right) +\left\vert g,1\right\rangle _{1}\left\vert
e,0\right\rangle _{2}\left\vert g,0\right\rangle _{3} \\
+\left\vert g,0\right\rangle _{1}\left\vert e,0\right\rangle _{2}\left\vert
g,1\right\rangle _{3}+\left( \left\vert g,1\right\rangle _{1}\left\vert
g,0\right\rangle _{2}+\left\vert g,0\right\rangle _{1}\left\vert
g,1\right\rangle _{2}\right) \left\vert e,0\right\rangle _{3}\} \\
\fl\left\vert \psi _{23}\right\rangle =\frac{1}{\sqrt{3}}\left( \left\vert
g,1\right\rangle _{1}\left\vert g,1\right\rangle _{2}\left\vert
g,0\right\rangle _{3}+\left\vert g,1\right\rangle _{1}\left\vert
g,0\right\rangle _{2}\left\vert g,1\right\rangle _{3}+\left\vert
g,0\right\rangle _{1}\left\vert g,1\right\rangle _{2}\left\vert
g,1\right\rangle _{3}\right) \\
\fl\left\vert \psi _{24}\right\rangle =\frac{1}{\sqrt{3}}\left( \left\vert
e,1\right\rangle _{1}\left\vert g,0\right\rangle _{2}\left\vert
g,0\right\rangle _{3}+\left\vert g,0\right\rangle _{1}\left\vert
e,1\right\rangle _{2}\left\vert g,0\right\rangle _{3}+\left\vert
g,0\right\rangle _{1}\left\vert g,0\right\rangle _{2}\left\vert
e,1\right\rangle _{3}\right) \\
\fl\left\vert \psi _{25}\right\rangle =\frac{1}{\sqrt{3}}\left( \left\vert
g,2\right\rangle _{1}\left\vert g,0\right\rangle _{2}\left\vert
g,0\right\rangle _{3}+\left\vert g,0\right\rangle _{1}\left\vert
g,2\right\rangle _{2}\left\vert g,0\right\rangle _{3}+\left\vert
g,0\right\rangle _{1}\left\vert g,0\right\rangle _{2}\left\vert
g,2\right\rangle _{3}\right)
\end{eqnarray*}
and same number of the solutions for characteristic equation for $g=\chi $ are $\lambda _{2,1}=-2.43065$, $\lambda _{2,2}=-0.771049$, $\lambda
_{2,3}=0.294764$, $\lambda _{2,4}=1.73598$, $\lambda _{2,5}=4.17096$. Here $\lambda _{i,j}=\lambda _{i,j}/g$. For three excitation, from 38 states we obtain 10 collective states. It is clear that for two or more excitations in three cavities we can solve the Schrodinger equation only numerically.

\textit{N cavity case.} Let us considerate that we have N cavities placed at the vertices of a regular polygon with N equal sides. If the number of cavities is large as this is represented in Fig.\ref{figure}, according to the interaction Hamiltonian (\ref{eq:HI}\}), for one undistinguished excitation of atoms or photons we observe the following degenerates wave functions
\begin{eqnarray*}
\left\vert \psi _{1}\right\rangle &=&\frac{1}{\sqrt{N}}\left( \left\vert
e_{1}g_{2}..g_{N}\right\rangle +...+\left\vert
g_{1}g_{2}..e_{N}\right\rangle \right) \left\vert
0_{1}0_{2}..0_{N}\right\rangle \\
\left\vert \psi _{2}\right\rangle &=&\frac{1}{\sqrt{N}}\left\vert
g_{1}g_{2}..g_{N}\right\rangle \left( \left\vert
1_{1}0_{2}..0_{N}\right\rangle +...+\left\vert
0_{1}0_{2}..1_{N}\right\rangle \right)
\end{eqnarray*}%
Is not difficult to observe that the actions of the interaction Hamiltonian on these functions are $H_{I}\left\vert \psi _{1}\right\rangle =\hbar g\left\vert \psi _{1}\right\rangle $ and $H_{I}\left\vert \psi _{2}\right\rangle =\hbar g\left\vert \psi _{1}\right\rangle +2\hbar \chi
\left\vert \psi _{2}\right\rangle $. In this case, the eigenvalues and eigenfunctions of interaction Hamiltonian are equivalent to the three
cavities case $\lambda _{1,1(-1)}=\chi \pm \sqrt{\chi ^{2}+g^{2}}$ and $c_{i,1}=\sqrt{g^{2}/(g^{2}+\lambda _{i}^{2})}$,$\ c_{i,2}=\sqrt{\chi
^{2}/(g^{2}+\lambda _{i}^{2})}\ $for$\ i=-1,1$.

In conclusion, we observe the following behavior of the excited atoms in two or more cavities. Two excitations in two cavities are described by 5 collective states. In three cavities, three excitations are described by 10 collective states. So we can observe that for small number of excitations can be applied the following expression for the number of the collective states $N=n_{ex}^{2}+1$, where $n_{ex}$ is the number of excitations in the system. This behavior is applicable for number of excitations $n_{ex}$ less than the number of cavities. This law is violated beginning with four excitations. Indeed for four excitation it is not difficult to observe that we can construct 19 collective states. According to our representation this corresponds to the number $n_{ex}^{2}+3$ . In our opinion this modification is connected with the packing of the cavities in the space, the atomic and photon excitations between cavities. For example four cavities can be placed in the circle and, of course,these cavities can be packed in the vertices of an regular tetrahedron. So we have two different types of symmetry, two cases with different number of collective states for four excitations.

\section{Master equation and behavior of population of new quasi-energy levels}
\label{Sec:3}
Let us obtain now the master equation for collective excitations of two radiators placed in two coupled cavities with the losses. In order to find the master equation for such system let us introduce in the Hamiltonian (\ref{eq:HI} ) the coupling of cavity modes with vacuum of EMF $H_{I}^{b}=\sum\limits_{i=1}^{n}\sum\limits_{k=1}^{m}\kappa a_{i}b_{k}^{\dagger }+H.c$. \ Here the photon losses $\kappa $ indicate the transformation of cavity photons in the free EMF photons described by the new annihilation and creation Boson operators $b_{k}$ and $b_{k}^{\dagger }$. Using the collective excitation states (\ref{eq:C1}, \ref{eq:C2})obtained in the first section we can represent the cavity field operators through the new collective states. Considering that the energy is measured from the relatively level $E=0$ (see Fig.\ref{figure}), the Hamiltonian for two coupled cavities can be represented as $H=\hat{H}_{0}+$ $\hat{H}_{I}$, where
\begin{eqnarray}
\hat{H}_{0} &=&\sum\limits_{i=-1,1}^{2}\hbar \tilde{\omega}\left\vert
1i\right\rangle \left\langle 1i\right\vert +\sum\limits_{k}\hbar \omega _{k}%
\hat{b}_{k}^{\dagger }\hat{b}_{k},  \nonumber \\
\hat{H}_{I} &=&\sum\limits_{k}\kappa \hat{b}_{k}^{\dagger
}\sum\limits_{i=-1,1}\{\sqrt{2}c_{i,2}^{(1)}\left\vert 0\right\rangle
\left\langle 1i\right\vert +H.c.\}+\hbar \sum\limits_{j=-1,1}j\lambda
_{1}\left\vert 1j\right\rangle \left\langle 1j\right\vert  \label{eq:HIM}
\end{eqnarray}
Here $\left\vert 1i\right\rangle =\sum\limits_{j=1,2}c_{i,j}^{\left(1\right) }\left\vert \psi _{1j}\right\rangle $ are the Hilbert vectors of
collective excitation, it is introduced re-normalized frequencies $\tilde{\omega}\pm \lambda _{1}=\omega +\lambda _{1,1(-1)}$, where $\tilde{\omega}=\omega +\frac{1}{2}\chi $ and $\lambda _{1}=\frac{1}{2}\sqrt{4g^{2}+\chi^{2}}$, $c_{1(2)}=c_{1(-1),2}^{(1)}$. In Hamiltonian we have considered that the action of the creation operators in the state with maximal number of excitations $m=1$ is neglected. $a_{i}^{\dagger }\left\vert \psi_{ml}\right\rangle =0$ due to the fact that such an excitation is impossible in the system in interaction with the external vacuum field.

Considering the large number of degrees of freedom of free EM vacuum, we can eliminate from the density matrix equation the boson operators of external EMF. In the interaction picture the density matrix equation $i\hbar\,\partial \check{\rho}(t)/\partial t=[\check{H}_{_{I}}(t),\check{\rho}(t)]$ where $\check{H}_{I}(t)=\exp [i\hat{H}_{0}t/\hbar ]\hat{H}_{I}\exp [-i\hat{H}_{0}t/\hbar ]$ is the Hamiltonian in interaction picture. Using the method of the projection operator on the vacuum field $\mathcal{P=}\left\vert 0\right\rangle \left\langle 0\right\vert Tr_{ph}\{..\}$ we can represent the density matrix through slower $\rho _{s}(t)=\mathcal{P}\check{\rho}(t)$ and
rapidly oscillating $\check{\rho}_{b}(t)=\bar{\mathcal{P}}\check{\rho}(t)$ parts respectively, $\mathcal{\overline{P}}=1-\mathcal{P}$. It can be shown that $\mathcal{P}^{2}=\mathcal{P}$ and $\mathcal{\overline{P}}\mathcal{P}=0$. Following the well known procedure of the elimination of the rapid oscillatory part of the density matrix \cite{C-1993} , one can obtain the equation of the density matrix $\check{W}(t)=Tr_{ph}\{\check{\rho}(t)\}$ for the cavity excitations
\begin{eqnarray}
\frac{d\check{W}\left( t\right) }{dt} &-&i\lambda _{1}\check{[W}\left(
t\right) ,U_{1,1}^{1,1}(t)-U_{1,-1}^{1,-1}(t)]  \nonumber \\
&=&-\gamma \lbrack \check{W}(t)\left( c_{1}\check{U}_{0}^{1,1}(t)+c_{2}%
\check{U}_{0}^{1,-1}(t)\right) ,c_{1}\check{U}_{1,1}^{0}(t)+c_{2}\check{U}%
_{1,-1}^{0}(t)]  \nonumber \\
&+&H.c.  \label{4}
\end{eqnarray}
Here $\gamma =\sum\limits_{k}$ $2\pi \delta \left( \omega _{k}-\tilde{\omega}\right) \kappa ^{2}/\hbar ^{2}$, $U_{\beta }^{\alpha }=\left\vert \alpha\right\rangle \left\langle \beta \right\vert $ is the Ladder operator which indicate the transition between the new quasi-levels for a single cavity-atom excitation. These operators satisfy the commutation relation $\left[ \check{U}_{\beta }^{\alpha },\check{U}_{\upsilon }^{\mu }\right]=\delta _{\beta ,\mu }\check{U}_{\upsilon }^{\alpha }-\delta _{\alpha,\upsilon }\check{U}_{\beta }^{\mu }$. The generalized equation (\ref{4}) takes into consideration the cooperative decay processes of two cavities in the interaction with the external vacuum field.

Let us study the decay rate of single collective excitation placed in the two or three cavities. Considering that the new collective states $%
\left\vert 1,1\right\rangle $ and $\left\vert 1,-1\right\rangle $ can be easily \ prepared, we found the following closed system of equations for the population and correlation of photons and radiators
\begin{eqnarray*}
\frac{dx(\tau )}{d\tau } &=&-x(\tau )-pu(\tau ), \\
\frac{dy(\tau )}{d\tau } &=&-p^{2}y(\tau )-pu(\tau ), \\
\frac{du(\tau )}{d\tau } &=&-2qw(\tau )-\left[ p^{2}+1\right] u(\tau )-\frac{%
p}{2}(x(\tau )+y(\tau )), \\
\frac{dw(\tau )}{d\tau } &=&2qu(\tau )-\left[ p^{2}+1\right] w(\tau ).
\end{eqnarray*}%
Here $\tau =t/\tau _{2}$, $p=c_{2}/c_{1}$, $q=\lambda _{1}/\tau _{2}$, $x=\left\langle U_{1,1}^{1,1}\right\rangle $, $y=\left\langle
U_{1,-1}^{1,-1}\right\rangle $, $u=\left( \left\langle U_{1,1}^{1,-1}\right\rangle +\left\langle U_{1,-1}^{1,1}\right\rangle
\right) /2$ and $w=\left( \left\langle U_{1,-1}^{1,1}\right\rangle -\left\langle U_{1,1}^{1,-1}\right\rangle \right) /2i$. For simplicity, we
have represent the relaxation times of two transitions from two upper to the ground states through the constants $1/\tau _{1(2)}=2\gamma c_{1(2)}^{2}$. The evolution of the population of new quasi-levels of energy is studied considering the situation when only the state $\left\vert 1,1\right\rangle $($\left\langle U_{1,1}^{1,1}(0)\right\rangle =1$) is populated while the second excited state $\left\vert 1,-1\right\rangle $ is un-populated. It is remarkable for physical point of view the partially transfer of excitation from the state $\left\vert 1,1\right\rangle $ to the state $\left\vert 1,-1\right\rangle $. This transfer is represented in the Fig. 2: in which figure $b.$represent the nutation of the excitation between the upper states $\left\vert 1,1\right\rangle $ and $\left\vert 1,-1\right\rangle $. This nutation is accompanied with the transfer of the excitation to the state $\left\vert 1,-1\right\rangle $ which in the processes of interaction with vacuum field becomes entangled with the state $\left\vert 1,1\right\rangle $. During the decay time, both states $\left\vert
1,1\right\rangle $ and $\left\vert 1,-1\right\rangle $ become unpopulated.
\begin{figure}[tbp]
\begin{centering}
\resizebox{0.34 \textwidth}{!}{\includegraphics{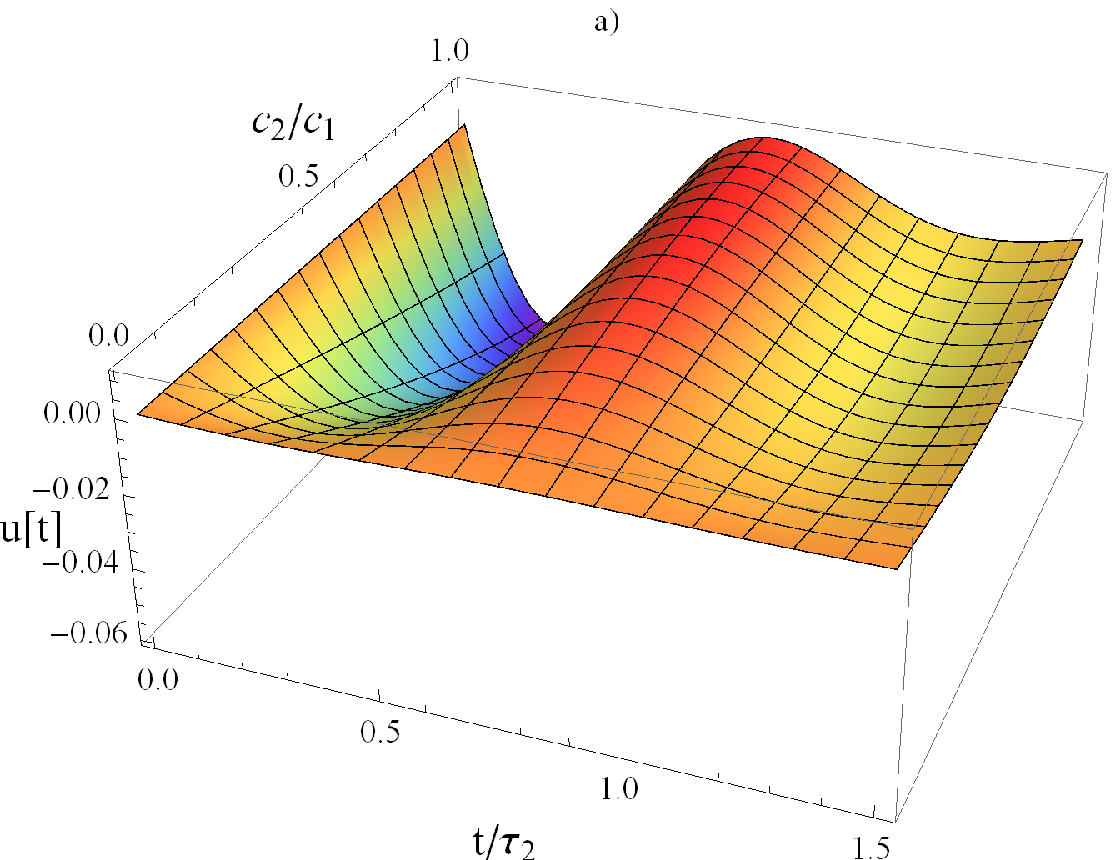}} \resizebox{0.35 \textwidth}{!}{\includegraphics{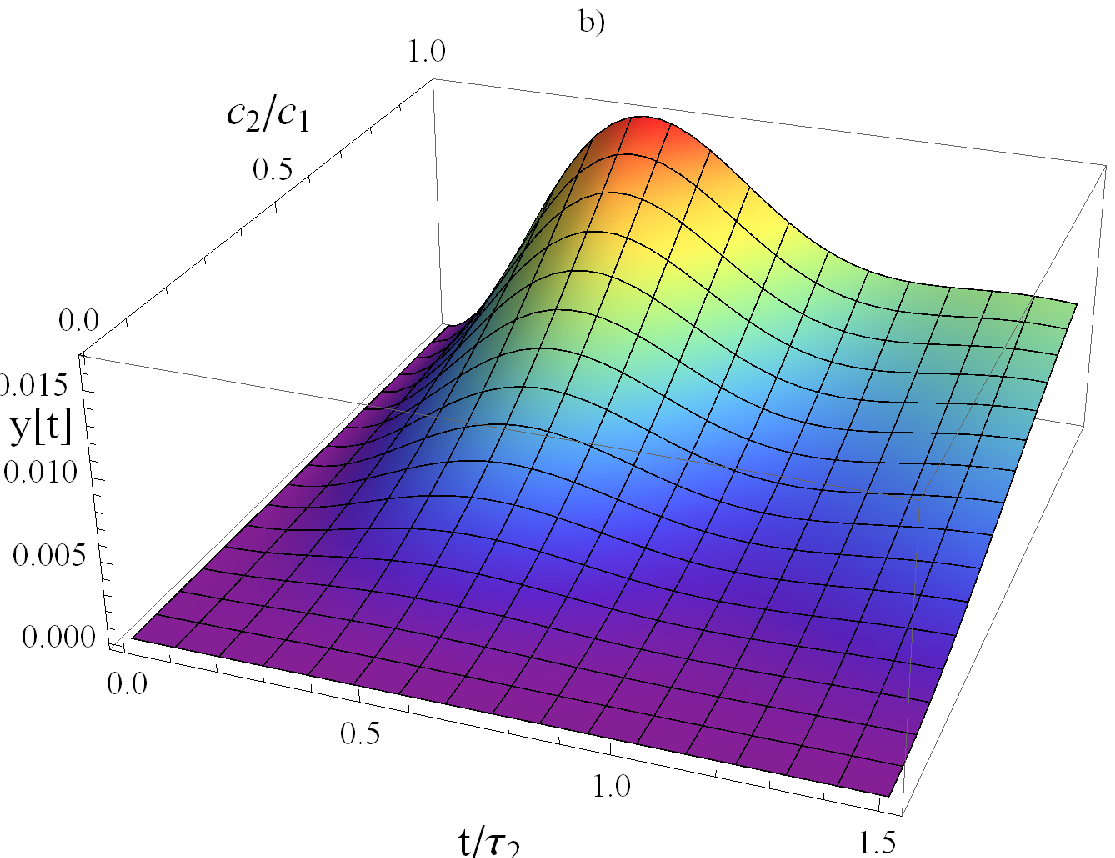}}
\caption{The relative time-dependence of population $y\left(\tau\right)=\left\langle U_{1-1}^{1-1}(\tau)\right\rangle $
of level $\left\vert 1,-1\right\rangle $ and of quantum nutation
$u(\tau)=$ $\left(\left\langle U_{1-1}^{11}(\tau)\right\rangle +\left\langle U_{11}^{1-1}(\tau)\right\rangle \right)/2$
for initial conditions $\left\langle U_{1-1}^{1-1}(0)\right\rangle =0$,
$\left\langle U_{11}^{11}(0)\right\rangle =1$, $q=3$.}
\par\end{centering}
\par
\label{figure_2}
\end{figure}
In order to obtain the entropy of excitations in the cavities let us represent the density matrix through the excitation correlations
\begin{eqnarray*}
\check{W}(t) &=&(1-x(t)+y(t))\check{U}_{0}^{0}+x(t)\check{U}_{1,1}^{1,1}+y(t)%
\check{U}_{1,-1}^{1,-1} \\
&&+(u(t)-iw(t))\check{U}_{1,-1}^{1,1}+(u(t)+iw(t))\check{U}_{1,1}^{1,-1}.
\end{eqnarray*}%
The entropy of the cavity excitations is described by the expression
\[
S=-\langle \check{U}_{0}^{0}(t)\rangle \log \langle \check{U}
_{0}^{0}(t)\rangle -U_{e1}\log U_{e1}-U_{e2}\log U_{e2}
\]
where $U_{e1}=(x+y)/2+\sqrt{(x-y)^{2}/4+w^{2}+u^{2}},$ $U_{e2}=(x+y)/2-\sqrt{(x-y)^{2}/4+w^{2}+u^{2}}$and $\langle \check{U}_{0}^{0}(t)\rangle =1-x-y$. From the time-dependence of entropy and inversion, we observe what then the population of excited levels becomes equal to the population of the ground state, the entropy takes the maxim value. This maximum corresponds to the creation of an entangled state between excited and ground states(see Fig.\ref{figure_4}). Another interesting effect is the decay process of excited states. Here is possible two situations, first of them corresponds to the preparation of excited states in the superposition of the states $\left\vert 1,1\right\rangle $ and $\left\vert 1,-1\right\rangle $. In this case the entropy become increases from non zero value $S(0)=0.5$. Preparing the system in only one excited state, we can observe, during the process of spontaneous emission that the excitation partially "jump" in the second excited state.  $\left\vert 1,1\right\rangle $ when the population of another state $ \left\langle U_{1,-1}^{1,-1}\left( 0\right) \right\rangle =0$. In this case, after the decay process of spontaneous emission, the electron jump on the unpopulated state. Similar effect is known in the literature and experimentally observed in three levels systems \cite{BRT-2004}.
\begin{figure}[tbp]
\begin{centering}
\resizebox{0.35\textwidth}{!}{\includegraphics{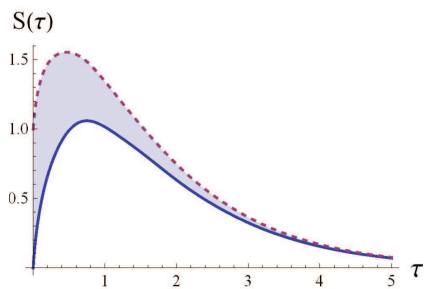}}
\caption{The relative time dependence of entropy for initial conditions $\left\langle \hat{U}_{1-1}^{1-1}(0)\right\rangle =0$, $\left\langle \hat{U}_{11}^{11}(0)\right\rangle =1$(continue line), $\left\langle\hat{U}_{1-1}^{1-1}(0)\right\rangle =1/2$, $\left\langle \hat{U}_{11}^{11}(0)\right\rangle =1/2$ (dashed line), $q=3$, $p=1$}
\label{figure_4}
\end{centering}
\end{figure}
\section{Conclusions}
In this paper we have introduced the collective excitations of EMF of coupled cavities in interaction with two level radiators placed in each
cavity (\ref{eq:3c}, \ref{eq:3cv}). Of course, with increasing the number of excitations, the problem becomes difficult in the analytical representation ( see two cavities with two excitations (\ref{eq:L2}, \ref{eq:C2})). Unlike the traditional approach Refs \cite{MO-2011,EB-2013} it examines the behavior of two collective excitations of coupled cavities taking into consideration local symmetry of this system, (see (\ref{eq:L2}) and (\ref{eq:C2})). Taking into consideration the undistinguished principle between the atomic and photon excitations this paper proposes to introduce the collective excitations for the description of the interaction q-bits. For this we proposed the exact solutions of two excitations in two cavities (\ref{eq:3c}) and (\ref{eq:3cv}). Considering that the energy of collective dressed states of atom +cavity field is larger than the decay rates from the cavities we have introduced the new interaction Hamiltonian of collective excitations with external vacuum field (\ref{eq:HIM}). From analytical and numerical solutions follows that the solution of a damped system of equations depends on the initial preparation of the collective states $\left\langle U_{1,1}^{1,1}(0)\right\rangle =\alpha_{1}^{2}$ and $\left\langle U_{1,-1}^{1,-1}(0)\right\rangle =\alpha_{2}^{2}$. In comparison with other approaches we observe the damped oscillations of the population
inversions between two excited states as it is represented in the Fig.\ref{figure_2}. It is observed two limits for this association. First limit $c_{2}/c_{1}\rightarrow1$ corresponds to the bad coupling between the cavities and second limit $c_{2}/c_{1}\rightarrow0$ is connected to high coupling between the cavities.

\end{document}